\documentclass[12pt,a4paper]{article}
\usepackage{amsthm}
\theoremstyle{plain}
\usepackage{authblk}
\usepackage[english]{babel}
\usepackage{graphicx}
\usepackage[latin1]{inputenc}
\usepackage{verbatim}
\usepackage{amsfonts}
\usepackage{amsmath}
\usepackage{txfonts}
\usepackage[T1]{fontenc}
\usepackage{color}
\usepackage{epsfig}
\usepackage{natbib}
\usepackage{url}
\usepackage[bookmarksnumbered=true, colorlinks=true, citecolor=blue]{hyperref}
\date{}

\title{Can Bohmian Mechanics Be Made Background Independent?}
\author{Antonio Vassallo}
\affil{University of Lausanne, Department of Philosophy, CH-1015 Lausanne\\ \url{antonio.vassallo@unil.ch}}

\begin{document}

\maketitle
\begin{center}
Forthcoming in \emph{Studies in History and Philosophy of Modern Physics}.
\end{center}
\pdfbookmark[1]{Abstract}{abstract}
\begin{abstract}
The paper presents an inquiry into the question regarding the compatibility of Bohmian mechanics, intended as a non-local theory of moving point-like particles, with background independence. This issue is worth being investigated because, if the Bohmian framework has to be of some help in developing new physics, it has to be compatible with the most well-established traits of modern physics, background independence being one of such traits. The paper highlights the fact that the notion of background independence in the context of spacetime physics is slippery and interpretation-laden. It is then suggested that the best-matching framework developed by Julian Barbour might provide a robust enough meaning of background independence. The structure of Bohmian dynamics is evaluated against this framework, reaching some intermediate results that speak in favor of the fact that Bohmian mechanics can be made background independent.\\

\end{abstract}

\textbf{Notation:} In the following, all equations will be written in natural units such that $c=G=\hbar=1$. The symbols $\mathbf{Eins}$, $\mathbf{Riem}$, $\mathbf{Weyl}$, $\mathbf{R}$ will designate, respectively, the Einstein, Riemann, and Weyl tensors, and the curvature scalar.

\section{Introduction}
It is widespread opinion among physicists and philosophers that modern physics rests on two robust theoretical pillars, namely, quantum theory and general relativity. The motivation for this opinion has several deeply intertwined aspects such as strong empirical corroboration, effectiveness in treating a wide range of concrete cases, and mathematical beauty. From a scientific realist perspective, the current situation might be summarized as follows: although neither quantum theory nor general relativity are ``final theories'', still they succeed in grasping some of the most fundamental facts regarding the physical world. It is then obvious that whatever further theoretical development - be it the quest for a final theory that unifies all the aspects of reality or, (not much) more modestly, the effort to construct a theory that provides a quantum description of gravitational phenomena - is very likely to preserve the main constitutive features of quantum theory and general relativity. The problem, however, is to clearly spell out what such constitutive features are and what physical significance they bear: this is where the widespread consensus breaks down and makes way for a lively debate on the interpretation of the afore-mentioned theories.\\
For example, one of the most debated interpretational problems - perhaps the interpretational problem par excellence in modern physics - is the measurement problem in quantum physics. \citet{431} effectively summarizes the core of the issue in the following question: <<Why do measurements have outcomes at all, and what selects a particular outcome among the different possibilities described by the quantum probability distribution?>> (\emph{ibid.}, p. 50).\footnote{To be fair, Schlosshauer claims that this is just an aspect - that he dubs ``the problem of outcomes'' - of the overall issue, which includes ``the problem of preferred basis'' and ``the problem of the nonobservability of interference''. However, the above characterization of the problem is sufficient for our purposes. The interested reader can also refer to \citet{197} for an extensive philosophical discussion of the measurement problem.} Such a question naturally arises once it is claimed that the complete physical description of any quantum system (including measuring apparatuses) is encoded in a wave function $\Psi$ (evolving in time according to the Schr\"odinger equation), and becomes all the more compelling when this claim is taken to entail that all there is to the world is just wave functions.\\
Many solutions to the measurement problem have been proposed over the years, some of them involving a mere reinterpretation of the standard formalism - perhaps on the ground of a brand new logic -, some others providing further theoretical structure. This paper will be concerned with one among these latter proposals, namely, the de Broglie-Bohm theory in the form set out in \citet{323}, commonly referred to as Bohmian mechanics (BM). To cut the story short, BM succeeds in solving the measurement problem (at least, in the non-relativistic case) by (i) postulating an ontology of material point-particles with definite positions in Euclidean 3-space at all times, and (ii) providing a deterministic law of motion depending on $\Psi$ for the temporal development of the universal configuration of such particles. In Bohmian terms, the measurement process is just a complicated ``dance'' involving the particles constituting the measured system and those making up the measuring apparatus. A measuring outcome - for example, a pointer pointing on a display, or a spot on a photographic plate - is thus just a certain configuration of particles as observed at a given time. In BM, there is no real indeterminism regarding the possible outcome of a measurement. By taking into account - besides $\Psi$ - also the initial positions of the particles, the law of motion singles out the final configuration of the system plus measuring apparatus, and hence a definite outcome.\footnote{For a detailed defense of the claim that BM solves the measurement problem, see \citet{233}.}\\
We will be more precise about the formal machinery of BM in the next section. For the time being, we just note that BM is also able to account for one of the most puzzling fundamental facts about reality that the quantum theory captures, that is, quantum non-locality. By this designation it is usually intended the empirically proven fact\footnote{See, for example, \citet{248,249}.} that the statistical distributions of outcomes of measurements performed on - to fix the ideas - two ensembles of quantum systems $\mathcal{E}(A)$ and $\mathcal{E}(B)$ prepared such that $A$ and $B$ are in an entangled state, are correlated no matter how far in space $A$ and $B$ are placed, such statistical correlation being not explainable in terms of a past common cause.\\
The fact that BM tells a coherent story about quantum measurement processes - including the explanation of the non-local character of some of them - makes it a serious contender for providing a robust interpretational framework for future developments in quantum theory. However, if BM has to prove as a really useful approach in developing new physics, it cannot ignore the second pillar of modern physics, namely, general relativity. In such a context, what a Bohmian approach should provide in order to be taken seriously by the scientific community is a way out from the apparent incompatibilities between quantum theory and relativity.\\
Just to have a hint of one of such incompatibilities, consider the embedding of non-locality in a relativistic spacetime, say the Minkowski one of special relativity. In this context, the above sketch of non-local correlations can be translated by saying that an event $E(A)$, such as the appearance of an outcome for a measurement performed on $A$, can be influenced by other events happening outside the past lightcone of $E(A)$. Such a conceptual tension between non-locality and the lightcone structure of relativistic spacetimes is inescapable for whatever theory that seeks to account for quantum phenomena, hence also for BM. The question of the compatibility between BM and relativity has been tackled in several papers;\footnote{See, for example, \citet{408}.} however, we have the impression that just one particular aspect of the problem has been extensively discussed so far.\\
To better frame our complaint, we need to agree on what the main tenets of (general) relativistic physics are, and this is surely a controversial matter. Anyway, we think that the most part of physicists and philosophers would agree at least on two points. First of all, in relativistic physics the only salient spatiotemporal structure is that encoded in a $4$-dimensional semi-Riemannian metric tensor. Here, with ``salient'' we mean that the most necessary physical information about a spatiotemporal setting is specified once and for all once a metric is given. This is, of course, not to say that the specification of a metric automatically exhausts the physical description of a spacetime (e.g. a metric alone does not fix a unique connection, so that we have to specify some extra condition, such as compatibility, in order to associate a single connection to a metric), 
but it is just to claim that, in a relativistic context, no geometrical object is needed over and above a single $4$- dimensional semi-Riemannian metric in order to describe the metrical properties of spacetime.\footnote{This is what \citet[][p. 292]{238} calls \emph{Relativistic Constraint}.} The second key requirement is the absence of spatiotemporal structures that act as backgrounds (we will call this feature \emph{background independence}). One of the most important lessons from general relativity, in fact, is that spacetime is not an inert arena in which physical interactions between material objects take place, but it is by itself a dynamical entity that interacts with matter. Both issues are overwhelmingly important in the general quest for the unification of the ``relativistic'' with the ``quantum''. As already mentioned, the spatiotemporal structure needed to accommodate quantum non-locality is very unlikely to be encoded just in a relativistic $4$-metric - into which the lightcone structure in fact inheres -, and background independence is also incompatible with quantum theory, because this latter theory is formulated over an inert ``container'', be it a Newtonian or Minkowski spacetime.\\
Returning to the special case of BM, in our opinion, what it is that has been discussed so far is the compatibility of BM with the former relativistic tenet, that is, that all of the structure inhering within spacetime is encoded in the $4$-metric, but not much has been said on the latter, namely, whether the formulation of BM can dispense with spatiotemporal backgrounds. Under the light of the above mentioned incompatibility within quantum theory and general relativity, this question is extremely interesting because, if it turned out that  the structure of the dynamics in BM cannot be made compatible with a background independent context, then we would have a strong argument against viewing the Bohmian approach as a useful framework in the development of new physics that overcomes the dichotomy between the ``quantum'' and the ``relativistic''.\\
The paper will start by giving a general overview of BM, will continue by discussing how to sharpen the above presented notion of background independence in spacetime physics, and will end by considering the possibility of extending BM to a background independent context under the light of the analysis carried out.

\section{The Bohmian Approach to Quantum Physics}\label{2}
The roots of BM go back to the work of Louis \citet{338, 220}, later revived by David \citet{187a,187b}. Here we will discuss the version of the theory put forward, for example, in \citet[][especially section 3]{222}.\\
The core dynamical feature of BM is adding a further equation parallel to the Schr\"odinger evolution for the wave function:
\begin{subequations}\label{bm}
\begin{equation}\label{sch}
i\frac{\partial\Psi(\mathbf{Q},t)}{\partial t}=\Bigg (-\sum_{i=1}^{N}\frac{\nabla_{i}^{2}}{2m_{i}}+V\Bigg )\Psi(\mathbf{Q},t);
\end{equation}
\begin{equation}\label{gui}
\frac{d\mathbf{Q}}{dt}=\mathbf{m}^{-1}Im\frac{\boldsymbol{\nabla}\Psi}{\Psi}(\mathbf{Q},t).
\end{equation}
\end{subequations}
$\mathbf{Q}=\left ( \begin{array}{l}
\mathbf{q}_{1}\\
\cdots\\
\mathbf{q}_{N}
\end{array} \right )$ is just a point in $\mathbb{R}^{3N}$, which is the configuration space of the theory so formulated, $\boldsymbol{\nabla}=\left ( \begin{array}{l}
\boldsymbol{\nabla}_{1}\\
\cdots\\
\boldsymbol{\nabla}_{N}
\end{array} \right )$ is the ``gradient vector'', and $\mathbf{m}$ is the $N\times N$ diagonal ``mass matrix'' $\{\delta_{ij}m_{i}\}$.\footnote{Note that equations (\ref{bm}) can be made compatible with whatever Riemannian structure definable on configuration space \citep[][section 2]{257}. The relevance of this fact will become apparent in section \ref{4}.} In short, (\ref{gui}) represents $N$ coupled equations of the form:
\begin{equation}\label{gui2}
\frac{d\mathbf{q}_{k}}{dt}=\frac{1}{m_{k}}Im\frac{\boldsymbol{\nabla}_{k}\Psi}{\Psi}(\mathbf{Q},t).
\end{equation}
Formally, (\ref{gui}) depicts a vector field on $\mathbb{R}^{3N}$ depending on $\Psi$, whose integral curves $\mathbf{Q}=\mathbf{Q}(t)$ can be intended as collections of $N$ continuous trajectories $\{\mathbf{q}_{i}=\mathbf{q}_{i}(t)\}_{i=1,\dots,N}$ stretching in a spacetime with topology $\mathbb{R}^{3}\times\mathbb{R}$. The dynamics encoded in (\ref{bm}) is deterministic: once provided a set of initial conditions $(\Psi_{0},\mathbf{Q}_{0})$ at a fixed time $t_{0}$, the dynamics singles out a unique dynamical evolution at earlier and later times. Moreover, it can be shown that (\ref{bm}) recovers the Born's rule of standard quantum mechanics, thus matching all the empirical predictions of this latter theory.\footnote{See, for example, \citet{222} for a detailed technical justification of this claim.}\\
The physical interpretation of BM is straightforward: the theory talks about $N$ massive spinless\footnote{However, the theory can be easily generalized in order to account for phenomena involving spin, as shown, for example, in \citet{380}.} point-like particles with definite positions $\mathbf{q}_{k}=(x_{k},y_{k},z_{k})$ in Euclidean $3$-space at all times; the wave function in this picture has the role of generating the vector field on the right-hand side of (\ref{gui}), and it is thus said to ``guide'' the motion of the particles. The peculiar feature of BM, what makes it a \emph{quantum} theory as opposed to a classical one is evident in (\ref{gui2}), the motion of a particle being instantaneously dependent on that of \emph{all} the other $N-1$ particles. This is how BM implements non-locality: by virtue of (\ref{gui}) being a non-local law. Furthermore, BM is a universal theory in primis since (\ref{bm}) describes the dynamics of all there is in the universe - i.e. particles. However, the theory provides a consistent procedure for defining sub-configurations of particles approximately behaving as isolated quantum systems guided by an ``effective'' sub-wave function. It is exactly thanks to this fact that BM accounts for ordinary quantum measurements.\footnote{See, again, \citet[][section 5]{222}.}\\
BM breaks with standard quantum mechanics in a striking way: the former theory admits a clear ontology - particles with definite positions in $3$-space at each time - and dispenses with the reality of quantum superpositions in that, at each time, there is just a well-defined configuration $\mathbf{Q}$ of concrete objects localized in $3$-dimensional Euclidean space. Such an ontology of classical objects accounts for non-locality through the entanglement relation among particles encoded in the wave function. Here we do not want to dig too deep into the metaphysical characterization of $\Psi$ in BM, suffices it to say that the theory does not require to look at the universal wave function as some sort of physical field, but rather accords to it a nomological role.\footnote{An extensive metaphysical discussion on this subject is carried out, e.g., in \citet{230}.} Instead, the important point to be highlighted is that, among the ontic commitments of the theory, there is not only that to particles, but also that to a Euclidean $3$-space and an objective universal time: these two latter elements enter (\ref{bm}) exactly in the same way as they enter in classical mechanics, namely, they are just an arena where particles move. In a word, BM as sketched here is a background dependent theory in a very clear and intuitive sense. Is there any possibility to modify (\ref{bm}) in order to render the theory background independent? Before attempting to answer this question, we need to clarify the possible meaning of background independence. In the next section, we will see why finding a definition of background independence, even guided by the clear Newtonian intuition of what a background is, is anything but a simple task.

\section{Background Independence: A Slippery Concept}\label{3}
The intuitive notion of background independence is, indeed, very simple: a theory is background independent if it is the case that the ``container-contained'' metaphor does not apply to it or, less metaphorically, if it is the case that all the ``actors'' entering the physical description are subjected in one way or another to physical interaction. However, when coming to a concrete characterization of background independence, this heuristic picture reveals itself to be rather feeble. This is mainly because there are many factors that enter the physical description provided by a theory, and it is not always simple to discriminate among them real ``actors'' from mere formal features.\\
Despite the conceptual difficulties that we are going to spell out in a moment, the requirement of background independence is of paramount importance in modern physics, and represents a constraint on any future theoretical development. There are many arguments - philosophical and physical - in favor of this view; here we limit to state the two most obvious among them. The first argument has a metaphysical flavor, and goes like this: a physical theory describes the physical interactions between physical entities, and the very notion of interaction requires some kind of reciprocal ``influence'' between the physical actors; it is thus suspicious to admit in the physical picture some entities - in fact, background structures - that influence others without being affected in return, since that would seem to imply the existence of strange one-way interactions. Absolute space and time in Newtonian mechanics are a perfect example of suspicious background structures: their existence and characterization is independent of the material content of the universe, yet they influence - indeed, they \emph{define} - the motion of material bodies. The second argument is more physical and refers to the second pillar mentioned in the first section, namely, general relativity. In this theory, the description of gravitational phenomena is indistinguishable from the characterization of spatiotemporal structures (e.g. tidal forces as geodesic deviation), and hence the dynamics of gravitational interactions presupposes a dynamical nature of spatiotemporal structures. In short, in general relativity space and time are not backgrounds in a Newtonian sense, since they are influenced by the material content of the universe. Therefore, any modern theory that seeks to incorporate gravitational phenomena, or that just claims to be compatible with their description, needs at least to be compatible with background independence.\\
Having clarified the motivation for looking at background independence as a very important component entering modern physical theories, let us continue by laying down a semi-formal sketch that can help us to carry out an analysis of background independence. Just to keep things simple, we agree to model spacetime as a $4$-dimensional differentiable manifold $M$ over which several geometric objects can be defined either as mappings from $M$ to another space $X$ (e.g. if $X$ is a vector space, then some properly defined map $\phi:M\rightarrow X$ would be a vector field over $M$) or, viceversa, as mappings from $X$ to $M$ (e.g. if $X=\mathbb{R}$, then a continuous and differentiable mapping $\sigma:\mathbb{R}\rightarrow M$ would represent a parametrized curve $\sigma (\lambda)$ on $M$). This way of formulating a theory is usually called \emph{intrinsic} because it does not rely on the notion of coordinate system in order to be laid down. Of course, all the above framework can be translated in a coordinate dependent language, where the geometrical objects are given in terms of components (e.g., matrix elements) in a given coordinate system. Usually, the geometric objects defined over $M$ pertain to two distinct categories, namely, (i) those taken to encode the \emph{physical} geometric structure of $M$ (such as metric tensor fields, connections, curvature tensors, and the like), and (ii) those taken to represent material stuff displaced over spacetime. In a so construed theoretical framework, the ``physical happenings'' are described by a set of equations $\mathfrak{E}$ relating the geometric objects defined over $M$. A solution of $\mathfrak{E}$ is called a \emph{model} $\mathfrak{M}$ of the theory, from which physically observable quantities are supposed to be extracted for purposes of empirical testing. Such observable quantities depend on the physical degrees of freedom of the theory, which are represented by those geometric objects in $\mathfrak{E}$ which are actually subjected to dynamical evolution. This latter point introduces a vital distinction for assessing the background independence of a theory, that is, the distinction between dynamical and absolute objects. In short, to say that an object is absolute is to say that it appears unchanged modulo diffeomorphisms (that is, bicontinuous and differentiable mappings from $M$ into itself\footnote{In general, a diffeomorphism can relate different differentiable manifolds. Here, the word ``diffeomorphism'' will designate just those from $M$ to itself.}) in all the models of the theory. A theory is then background independent if it does not admit absolute objects.\footnote{This characterization was firstly developed by \citet{66} and further refined by \citet{15}.} Such a definition seems to nicely fit into the intuitive characterization provided at the beginning of this section. According to this criterion, Newtonian mechanics easily qualifies as background dependent, since all its models always admit the same geometrical structures, which characterize a Newtonian spacetime (e.g. distinguished spatial and temporal metrics). On the other hand, general relativity qualifies as a background independent theory: this is because a generic model of the theory has the form $\mathfrak{M}=<M,\mathbf{g},\mathbf{T}>$, the metric tensor field $\mathbf{g}$ and the stress-energy tensor $\mathbf{T}$ being both dynamical objects related by the Einstein's field equations $\mathfrak{E}\equiv \mathbf{Eins}[\mathbf{g}]=8\pi\mathbf{T}$. This means that, in general, taken two models of $\mathfrak{E}$, $\mathfrak{M}_{1}=<M,\mathbf{g_{1}},\mathbf{T_{1}}>$ and $\mathfrak{M}_{2}=<M,\mathbf{g_{2}},\mathbf{T_{2}}>$, it will be the case that $\mathbf{g_{1}}\neq \mathbf{g_{2}}$ and $\mathbf{T_{1}}\neq \mathbf{T_{2}}$. At this point, it seems that we have given a well-defined formal criterion for assessing the background independence (or lack thereof) of a theory: we just scan the solutions space of $\mathfrak{E}$ in search of ``persistent'' structures, and if we find them we flag them as absolute objects, thus judging the theory as background dependent; otherwise we conclude that the theory is background independent. Unfortunately, such formal criterion works just in extremely simple and clear cases.\\
An evident flaw in the above criterion is given by the fact that, according to it, even allegedly background independent theories might turn out to contain absolute objects. For example, in a general relativistic theory describing an universe filled with ``dust'', i.e. a perfect fluid with positive mass density and identically null pressure, the velocity vector field of this dust will be the same up to diffeomorphisms in all models of the theory, thus qualifying as an absolute object.\footnote{See \cite{354} for a detailed discussion of this and similar cases.} This example points out that the condition of being dynamical is not always sufficient to exclude absoluteness, and this is due to the fact that we can always render absolute objects dynamical by adding further conditions to $\mathfrak{E}$. Just to have a rough idea of why this is so, imagine we have a simple theory including a flat background spacetime represented by a Minkowski metric $\boldsymbol{\eta}$; in order to render such a theory background independent in the above sense, we just need to reformulate it in terms of a generic tensor field $\mathbf{g}$ that satisfies the dynamical condition $\mathbf{Riem}[\mathbf{g}]=0$, that is, which has no Riemannian curvature, i.e. it is flat. Obviously, in this very simple example, all solutions of the theory would feature a $\mathbf{g}$ which is just a diffeomorphic image of $\boldsymbol{\eta}$, thus making very easy to individuate the background in disguise;  however, the more complicated the theory, the less obvious is spotting possible absolute objects rendered dynamical in this way.\footnote{See \citet{47} for a self-contained yet extremely illuminating discussion of this type of issues in spacetime physics.} The reason why individuating background structures uncontroversially is not always simple is that, in general, deciding whether a structure is absolute or not is a matter of interpretation. Let us focus on this latter point by considering a theory describing the non-linear propagation of a scalar field $\phi$ over Minkowski spacetime. In this case, $\mathfrak{E}$ can be written:
\begin{subequations}\label{chenex2}
\begin{equation}\label{minchioski2}
\mathbf{Riem}[\mathbf{g}]=0,
\end{equation}
\begin{equation}\label{mesorotto2}
\boldsymbol{\Box}_{\mathbf{g}}\phi=-4\pi\phi^{3}\mathbf{T},
\end{equation}
\end{subequations}
which seems to clearly represent a background dependent theory, $\mathbf{g}$ being the background Minkowski metric ``camouflaged'' as a dynamical object. However, let us now consider a generic conformally flat metric $\mathbf{g'}$, such that:
\begin{subequations}\label{nor}
\begin{equation}\label{nor1}
\mathbf{g'}=\phi^{2}\mathbf{g},
\end{equation}
\begin{equation}\label{nor2}
\phi=(-det|g'|)^{\frac{1}{8}},
\end{equation}
\begin{equation}\label{nor3}
\mathbf{g}=\mathbf{g'}(-det|g'|)^{\frac{1}{4}},
\end{equation}
\end{subequations}
where $det|g'|$ can be seen as the determinant of a matrix representing the tensor $\mathbf{g'}$ in a given coordinate system. If we rewrite equations (\ref{chenex2}) using the metric $\mathbf{g'}$ so defined, we obtain:
\begin{subequations}\label{chenex3}
\begin{equation}\label{minchioski3}
\mathbf{Weyl}[\mathbf{g'}]=0,
\end{equation}
\begin{equation}\label{mesorotto3}
\mathbf{R}[\mathbf{g'}]=24\pi\mathbf{T}.
\end{equation}
\end{subequations}
Now, the theory (\ref{chenex3}) qualifies as background independent: for any two models $\mathfrak{M}_{1}=<M,\mathbf{g'_{1}},\mathbf{T_{1}}>$ and $\mathfrak{M}_{2}=<M,\mathbf{g'_{2}},\mathbf{T_{2}}>$, in general $\mathbf{g'_{1}}\neq \mathbf{g'_{2}}$ and $\mathbf{T_{1}}\neq \mathbf{T_{2}}$. It seems then, that we have reached a paradoxical situation: (\ref{chenex2}) and (\ref{chenex3}) \emph{given} (\ref{nor}) are equivalent, but while the former theory is background dependent, the latter is background independent. The way out of this unwanted situation is to recognize that the physical significance of (\ref{nor}) rests on the interpretation chosen. If we interpret the theory as describing a field propagating over Minkowski spacetime, then (\ref{nor}) is just a way to compactly write the couple $(\mathbf{g},\phi)$ using a geometric portmanteau $\mathbf{g'}$ which, however, is not to be intended as a ``real'' metric tensor field. If, on the other hand, we take the theory as describing a conformally flat dynamical metric tensor field, then (\ref{nor}) is a way to reformulate the framework on a flat fixed space by using a ``fake'' scalar field $\phi$.\\
At the moment, there are on the table different proposals for making sense of the notion of background independence even in these complicated cases. A remarkable analysis is developed by \citet{311}\footnote{From which the above example is taken.} in the context of field theories: the core of Belot's proposal is to think of background independence not as a all-or-nothing affair - as the above discussion centered on the notion of absolute object might suggest, but rather as a feature that comes in degrees. Also in this framework, background independence has not the status of a formal definition, but depends on the interpretation given to a theory.\\
There are two major morals to be drawn from the above quick overview that are useful for us. Firstly, spotting background structures in a theory is an interpretation-laden task. Secondly, claiming that a theory is background independent because all the spatiotemporal structures figuring in it are dynamical is not sufficient. The immediate consequence of these two facts is that there is no ``formal test'' that judges if a theory is hopelessly background dependent or might be rendered background independent in a non-trivial physical sense with some modifications in its formalism.
This conclusion is all the more important in our case, since we are trying to assess the compatibility between background independence and the Bohmian dynamics encoded in (\ref{bm}). 

\section{Background Independent Bohmian Dynamics}\label{4}
The discussion carried out in the previous section helps us evaluating the solidity of the conceptual ground on which we are to base our discussion on the compatibility between background independence and Bohmian dynamics. As it will become clear in a moment, this ground is fortunately firm enough to let us derive some intermediate results that can be spent in a research effort towards the implementation of a background independent Bohmian framework.\\
First of all, let us make extremely clear what kind of theory we are discussing here: the subject of our inquiry is just the theory (\ref{bm}), that is, a non-local theory of moving point-like particles simpliciter. It is important to point out this fact because there are some proposals on the table about a Bohmian theory where what is guided is the general relativistic gravitational field \citep[one of the most worked out among them being that proposed by][which treats the components of a Riemannian $3$-metric as the ``stuff'' to be guided]{161}. These theories try to implement background independence by construction, in the sense of spatiotemporal structures being dynamical ab initio. However, in such cases, considerations about background independence are deeply entangled with further (huge) conceptual issues rooted in the far more general attempt to quantize gravity. A decent discussion of this topic would be too long and technical, and hence cannot be put forward in this paper. Saving such concerns for future investigations, we note en passant that the preliminary results we will state here might be useful also in the quest for a Bohmian theory of quantum gravity.\\
So let us focus on (\ref{bm}), and firstly note that the way the equations are written does not satisfy the characterization of a physical theory adopted in the previous section. This is because, formally speaking, the equations are given in a coordinate dependent language and, furthermore, they are written in a particular coordinate system. Hence, they should at least be written in a general coordinate system in order to be translated in the intrinsic language involving geometrical objects defined over a manifold $M$. However, it is a simple exercise to show that the problem is solved by including in the theory some properly defined objects, such as a (co)vector field $\mathbf{t}$ that acts as a ``universal time'' for a stack of Euclidean $3$-spaces (planes of absolute simultaneity), each with a given $3$-metric tensor: in this way, we can for example substitute the ordinary time derivative with a covariant temporal derivative with components $t^{\mu}\nabla_{\mu}$ in a generic coordinate system $\{x^{\mu}\}$. With these adjustments in place, (\ref{bm}) can be rewritten in a way that satisfies our starting definition, thus allowing us to extend the considerations made in the previous section to the present case. After this task has been carried out,\footnote{The technical details should not bother us in the present context, see \citet[][chapter III]{15} for a comprehensive treatment of the intrinsic formulation of classical mechanics.} we immediately notice that the geometric objects introduced to generalize the theory are nothing but the spatiotemporal structures involved in the intrinsic formulation of classical mechanics. This comes at no surprise since, by construction, BM involves the picture of a dynamical evolution that physically unfolds in a neo-Newtonian or Galilean spacetime.\footnote{But not everyone agrees on this: someone claims that the dynamics of BM is better understood as taking place in a full Newtonian spacetime, where there is a notion of absolute rest \citep[see][in this respect]{409}. However, this disagreement is tangential to our analysis, since both neo- and full Newtonian spacetimes admit the same purely spatial symmetries, on which we will shortly focus. For this reason, from now on, we will obliterate the distinction between these two spacetimes.} This also makes evident a fact that was already clear at the beginning of the discussion: BM is a background dependent theory in a fairly clear sense, because the motion of the particles takes place over a Newtonian spacetime that affects their motion without being affected in return.\footnote{To be precise, BM is formulated in configuration space and not in spacetime. However, from what has been presented in section \ref{2}, it is straightforward to argue that the theory treats only the latter as the real physical space (this is not to say, of course, that there is no possibility at all to argue that the configuration space is in fact the real space; however, this marvellous point will not be considered here).} But if we talk about one-way interactions, there is an even more striking background structure appearing in BM, namely, the wave function. It is in fact evident from (\ref{bm}) that $\Psi$ influences the motion of the particles, but its evolution is not in turn affected by them. Hence, how can we even just dream to obtain a background independent Bohmian dynamics, given that the wave function is an indispensable feature of standard BM? Replying that, in fact, $\Psi$ is subjected itself to dynamical evolution is clearly not enough: if the above mentioned ``action-reaction'' requirement represents an effective rule of thumb to spot background structures in a theory, then claiming that $\Psi$ is a dynamical object because it obeys (\ref{sch}) has no more force than claiming that a semi-Riemannian $4$-metric satisfying $\mathbf{Riem}[\mathbf{g}]=0$ is not a background Minkowski metric. Fortunately, this worry can be easily put aside by noticing that $\Psi$ qualifies as a background structure (in the ``action-but-not-reaction'' sense) just in case we reify it, i.e. we consider it as a concrete entity (say, a field in configuration space) that literally pushes the configuration of particles along a trajectory. However, the power of formulating BM as (\ref{bm}) resides exactly in providing a straightforward interpretation of $\Psi$ not as an object, but as a law-like element of the formalism. To have a rough idea of why it is so, consider that the role of $\Psi$ in the present formulation of BM is to generate through (\ref{gui}) a vector field in configuration space whose integral curves are the physically possible dynamical evolutions of the system under scrutiny. This is analogous to that what happens in Hamiltonian mechanics: there we have a law-like object, the Hamiltonian function, which generates through the Hamilton's equations a vector field in phase space, whose integral curves are, again, the physically possible evolutions of a system. With this analogy in mind, we perfectly see why $\Psi$ fails to qualify as a background structure according to the ``action-reaction'' rule of thumb: in reality, it does not push the particles in a literal sense, so the question whether the particles react on it is meaningless, exactly as it is meaningless to ask whether a mechanical system reacts on the Hamiltonian.\footnote{This point is also discussed in \citet[][section 12]{229}.} Having clarified that it is the Newtonian background that makes BM a background dependent theory in an interesting sense, we can ask the next question, that is: Can BM be made background independent in a physically non-trivial sense? The discussion developed in section \ref{3} seems to lead towards a discouraging answer: of course we can render dynamical the spatial and temporal metrics of the theory, together with all the other spatiotemporal structures needed to formulate the dynamics (\ref{bm}), by adding further dynamical conditions, but the resulting theory could hardly be considered something more than an elaborate mathematical trickery. However, not all hope is lost yet: perhaps some alternative strategy can be proposed in order to strip BM of its Newtonian background in a substantial sense while retaining all the good physics in it. Fortunately, such a strategy exists and, roughly speaking, consists in showing that the Newtonian background is not a fundamental structure of the theory but ``emerges'' from an underlying framework that dispenses with it. Let us now turn to the details of such a strategy, originally proposed by Julian Barbour and his collaborators.\footnote{See \citet{83} for one of the first papers on the subject, and \citet[][especially sections 6 and 7]{421} for a philosophical appraisal of Barbour's strategy.}\\
For the time being, let us put aside BM and consider just a Newtonian theory of $N$ particles with a well defined total kinetic energy $T_{class}=\frac{1}{2}\sum_{i=1}^{N}m_{i}\frac{d\mathbf{q}_{i}}{dt}\frac{d\mathbf{q}_{i}}{dt}$ and interacting via a classical potential $V=V(\mathbf{q}_{1},\dots,\mathbf{q}_{N})$ being a pre-assigned function of the coordinates. Such a theory, like BM, is most perspicuously formulated in the configuration space $\mathbb{R}^{3N}$; hence these two theories, although being radically different from a dynamical perspective, ``share'' the same Newtonian background, namely, a spatiotemporal structure of the form $\mathbb{R}^{3}\times\mathbb{R}$ which, roughly speaking, amounts to a pile of Euclidean $3$-planes of absolute simultaneity glued together by a rigid time-like flow. According to Barbour, stripping absolute space and time from this theory amounts to constructing a physically equivalent theory that treats spatial degrees of freedom as gauge, and (similarly) regards Newtonian time $t$ as just one among many possible - and physically equivalent - choices of parametrization for the dynamical evolution of the system.\\
As regards the elimination of the spatial structure, the preliminary step to be taken is inquiring into the ways in which this background affects the physical characterization of a configuration of particles. To this extent, consider, just to fix the ideas, two possible configurations of $N=3$ particles $\mathbf{Q}_{1}$ and $\mathbf{Q}_{2}$ characterized by the fact that all the three particles in each of them have the same relative Euclidean distance: in short, the shape of both $\mathbf{Q}_{1}$ and $\mathbf{Q}_{2}$ is that of an equilateral triangle. Obviously, what distinguishes $\mathbf{Q}_{1}$ from $\mathbf{Q}_{2}$ is the position of the particles involved with respect to Euclidean $3$-space. More precisely, $\mathbf{Q}_{1}$ and $\mathbf{Q}_{2}$ represent two different embeddings of the same shape in Euclidean space. The notion of sameness involved here is the following: if we take the two configurations, we hold one fixed and we move the second in $3$-space just using transformations pertaining to the Euclidean symmetry group $E(3)$ - such as rigid translations and rotations - until it is ``juxtaposed'' with the starting one, then the two configurations overlap: they are perfectly \emph{matched}. The fact that the shape of the two configurations is preserved under different embeddings is due to the symmetries of Euclidean space. Intuitively, since Euclidean space is homogeneous (there is no distinguished place) and isotropic (there is no distinguished direction), we can place the triangular configuration under scrutiny wherever and with whatever orientation, without changing its shape. This reasoning can be extended to dilations, in the sense that we can include in the matching moves also the stretching of the distances between the particles in the configuration under the condition that the relative angles are preserved: in this case, we should enlarge the symmetry group to contain also uniform scalings, thus ending up with the \emph{similarity} group $Sim(3)$ of Euclidean $3$-space. We have now reached a crucial point: if we take as fundamental the notion of sameness provided by the matching moves, then we conclude that different embeddings of the same shape are just gauge degrees of freedom, since applying whatever transformation in $Sim(3)$ to a shape does not change it. In other words, we pass from standard Newtonian mechanics to a new theory that considers all spatial (or, better, embedding-related) differences as not fundamental. This, in turn, hints at the fact that, in this new theory, the spatial background is not fundamental either. In other words, we are willing to claim that the Newtonian background is not a core feature of our new theory.\\
In order to make this claim more precise, let us consider again the configuration space $\mathbb{R}^{3N}$; according to what we have said so far, this space contains redundancies consisting in all the points corresponding to configurations that have the same shape. Therefore, it is natural to take the fundamental configuration space of the theory as the one we get by quotienting out all these redundancies form the picture. In the present case, such fundamental space of shapes is $\mathcal{Q}_{0}=\mathbb{R}^{3N}/Sim(3)$. Under this new picture, the now ``emergent'' configuration space $\mathbb{R}^{3N}$ has a principal fiber bundle structure $\mathcal{Q}_{0}\times Sim(3)$, $\mathcal{Q}_{0}$ being the base space and the fibers $Sim(3)$ being the equivalence classes of shapes under Euclidean symmetry group transformations.\\
So much for the elimination of spatial degrees of freedom. The next step is to implement on the new fundamental configuration space $\mathcal{Q}_{0}$ a dynamics that does not depend on a distinguished ``temporal'' parameter. This is achieved by setting up a geodesic principle on $\mathcal{Q}_{0}$ analogous to Jacobi's principle in classical mechanics.\footnote{The standard reference for a formal introduction to this important variational principle of mechanics is \citet[][chapter V, section 6]{414}.} The basic idea is to define a modified Jacobi's principle - let us call it best-matching principle - that selects kinematically allowed shapes and matches them in such a way that their sequence represents a geodesic curve of $\mathcal{Q}_{0}$.\\
In order to specify when two different shapes are best matched, we need to define a ``distance'' between shapes. To this extent,\footnote{The technical details of the following procedure can be found, e.g., in \citet[][chapter 2]{413}.} we start by defining the action of $Sim(3)$ on a generic configuration $\mathbf{Q}_{i}$ as $\mathcal{S}(\boldsymbol{\omega},\boldsymbol{\epsilon})^{j}_{i}\mathbf{Q}_{j}$, with $\boldsymbol{\omega}$ a set of auxiliary fields that parametrizes the symmetries, and $\boldsymbol{\epsilon}$ the generators of these symmetries: this action represents ``vertical'' shifts along the fibers.
Secondly, we need to construct a covariant derivative $\mathbf{D}$ depending on $\boldsymbol{\omega}$ and $\boldsymbol{\epsilon}$ that measures ``horizontal'' change on a trial curve between neighboring fibers. With this machinery in place, it is possible to define a Jacobi-type action $S_{J}$ depending on the ``shifted'' particles' coordinates $\mathbf{q}_{i}=\mathcal{S}(\boldsymbol{\omega},\boldsymbol{\epsilon})^{j}_{i}\mathbf{q}_{j}$ as well as on their covariant derivatives, which, after some mathematical manipulations, takes the form:
\begin{equation}\label{jac}
S_{J}=\int d\lambda\sqrt{\mathcal{C}T},
\end{equation}
where 
\begin{equation}\label{tpar}
T=T_{class}\Bigg(\frac{dt}{d\lambda}\Bigg)^{2}=\frac{1}{2}\sum_{i=1}^{N}m_{i}\frac{d\mathbf{q}_{i}}{d\lambda}\frac{d\mathbf{q}_{i}}{d\lambda}
\end{equation}
is the parametrized kinetic energy written in terms of the shifted coordinates, and $\mathcal{C}=E-V$. $V$ is the classical potential, while $E$ is a quantity whose physical meaning will become clear in a moment. It is important to note that (\ref{jac}) is invariant under reparametrizations, that is, the action does not change whatever monotonically increasing parameter $\lambda\in\mathbb{R}$ we choose; hence, the coordinates $\mathbf{q}_{i}$ are not anymore parametrized by a distinguished time $t$. The best-matching principle is finally implemented by properly extremizing this action with respect to the auxiliary fields (i.e. by computing the variation $\delta S_{J}=0$).\\
The condition $\delta S_{J}=0$, although formally defined over $\mathbb{R}^{3N}$, suffices to define the geodesic principle on $\mathcal{Q}_{0}$ searched for. In fact, it features a flat metric on $\mathcal{Q}_{0}$ proportional to $T$, which is curved by a conformal factor $\mathcal{C}$ (we recall that all these objects are given in terms of shifted coordinates $\mathcal{S}^{*}\mathbf{q}$ and covariant derivatives $\mathbf{Dq}$). Extremizing the action amounts to finding the curves in $\mathcal{Q}_{0}$ that put together or ``stack'' the shapes having the minimal best-matched ``distance''.\footnote{An important physical consequence of implementing the best-matching procedure is the appearance of constraints on the canonical momenta that amount to the vanishing of the total momentum, the total angular momentum, and the total ``dilational'' momentum of the system in the center-of-mass rest frame. For more technical details, see \citet[][section 4]{419}.} 
Note that all this procedure does not involve any consideration based on ordinary spatial or temporal notions.\\
By varying the action (\ref{jac}) with respect to the auxiliary fields $\boldsymbol{\omega}$, we obtain the equations of motion, which for the $k$-th particle read:
\begin{equation}\label{mot}
\frac{d}{d\lambda}\Big ( \sqrt{\mathcal{C}T^{-1}}m_{k}\frac{d\mathbf{q}_{k}}{d\lambda}\Big )=\sqrt{T\mathcal{C}^{-1}}\boldsymbol{\nabla}_{k}\mathcal{C}.
\end{equation}
As expected,  (\ref{mot}) are given in terms of a freely specifiable monotonically increasing parameter $\lambda$, which is consistent with the claim put forward earlier that the dynamics of this theory dispenses with a distinguished temporal metric.\\
The Newtonian theory ``emerges'' from (\ref{mot}) once we fix a specific parameter $t$, such that:
\begin{equation}\label{gnutfix}
\frac{dt}{d\lambda}=\sqrt{T\mathcal{C}^{-1}}.
\end{equation}
If we substitute condition (\ref{gnutfix}) in (\ref{tpar}), we get $E=T_{class}+V$. This means that $E$ can be interpreted as the (fixed) energy of the system of particles under the parameter fixing  (\ref{gnutfix}). Furthermore, under condition (\ref{gnutfix}), (\ref{mot}) reduces to:
\begin{equation}\label{gnutto}
m_{k}\frac{d^{2}\mathbf{q}_{k}}{dt^{2}}=-\boldsymbol{\nabla}_{k}V,
\end{equation}
which is nothing but Newton's equations of motion for the $k$-th particle.\\
The overall picture of the best-matching theory proposed here is now clear. To recap: the theory is formulated over a Riemannian space $\mathcal{Q}_{0}$, with a Riemannian structure consisting of a flat ``kinetic'' metric $T$ curved by a conformal factor $C$. Extremizing the action (\ref{jac}) singles out the geodesics of this space, which can be arbitrarily parametrized by monotonically increasing real parameters. Note that, up to this point, all the quantities at play are mere geometrical objects. Things become interesting once the distinguished parameter $t$ is selected by (\ref{gnutfix}): in this case we recover (i) a relation that can be interpreted as the energy conservation theorem for a (universal)\footnote{It is quite obvious why we need a universal perspective: if we had many subsystems of particles, and we applied a distinct condition (\ref{gnutfix}) to each of them, then we would have no guarantee that the different $t$'s would ``march in step''.} system of $N$ particles, and (ii) the usual equations of motion for this system. In short, we have shown that a Newtonian theory of $N$ particles is ``emergent'' in a clear formal sense from a fundamental theory that does not discriminate among either different spatial embeddings of a configuration or the particular parametrization under which the dynamics unfolds. The clear formal sense mentioned above is, in fact, one closely related to gauge fixing: we start with a purely relational series of shapes in $\mathcal{Q}_{0}$ constituting a geodesic for (\ref{jac}), and then we fix a parameter $t$ (that is, we \emph{define} Newtonian time) thanks to which we ``lift'' such a curve to the class of curves that satisfy (\ref{gnutto}). In short, the presented framework represents a procedure to mathematically reduce the Newtonian theory (\ref{gnutto}) to the embedding-and-parametrization independent theory (\ref{mot}).\\
Here, of course, we have glossed over many important technical and conceptual aspects of the framework. For example, the original aim of Barbour and his collaborators is to show that a theoretical framework can be developed that implements Machian ideas, such as the relational nature of inertial effects, or the fact that the dynamics has to be given only in terms of observable quantities such as (ratios of) relative distances and angles between physical bodies.\footnote{For an accessible but still quite technical introduction to Barbour's framework, see \citet{390}.} However, what is important for our purposes here is the fact that a fairly well-defined theoretical framework exists that strips the Newtonian dynamics of a universal configuration of particles of its absolute spatiotemporal structures and provides a clear mechanism for the emergence of such background from a fundamentally background independent theory. \citet{410} in fact proposes to evaluate the background dependence of a theory exactly under the light of this framework: a theory is background dependent just in case it accords physical significance to the ``motion'' along the fibers defined by the group of spatiotemporal symmetries of the theory ($Sim(3)$, in the present case). In this sense, the procedure to eliminate such physical significance and turn it into simple gauge freedom by implementing a variational principle based on (\ref{jac}) surely represents a more cogent way to render a theory background independent than the usual way of making absolute objects dynamical.\\ 
Having assessed the above framework as useful and viable, it is now time to discuss whether we can apply it in the context of BM: if it turns out that this is the case, then we have found a very encouraging answer to our initial question regarding the compatibility between Bohmian dynamics and background independence.\\
A possible setting of the problem might be: can we quantize the theory (\ref{mot}) and then provide a guiding equation depending on the background independent wave function such that it singles out a  set of geodesics on the base configuration space $\mathcal{Q}_{0}$? This is indeed a reasonable strategy but it should face two big challenges. First of all, the quantization of reparametrization invariant systems is still not completely understood. For some authors, a consistent quantization of this kind of systems\footnote{Namely, reparametrization invariant systems, irrespective of the fact that they are particle or field systems.} should inevitably lead to a timeless Wheeler-DeWitt equation (with all the related well-known conceptual issues); however, people like \citet{405,430} challenge this conviction based on a rather technical analysis of the phase space quantization of these systems. Secondly, even if under a Bohmian perspective the fact that the wave function obeys a Wheeler-DeWitt-like equation is not too big a problem and, perhaps, it is even a welcome result,\footnote{In a Bohmian framework, thanks to the guiding equation, even a timeless Wheeler-DeWitt-like equation generates a non-trivial dynamics. Furthermore, since the wave function has a law-like status, the fact that it does not change in time but it is somehow selected once and for all ab initio by the Wheeler-DeWitt equation is perhaps more conceptually comfortable.} it is far from clear how to implement a guiding equation on $\mathcal{Q}_{0}$ such that, taking the background independent wave function as input, it generates a vector field whose integral curves are geodesics according to the best-matching principle. Although these conceptual difficulties are not a priori insurmountable, they show that this route is still so much little followed that it gives us not enough results on which we could base even a preliminary partial evaluation. Therefore, we now turn to another possible strategy that, albeit not yet implemented (to our knowledge), is more delineated to let us deliver a useful conceptual analysis. The basic idea underlying this second strategy is brutally simple: if it is too difficult to quantize a background independent theory, try to strip the background from an already quantized one.\\
Let us start by summarizing the best-matching procedure discussed above. First we started with a background dependent particle theory and we individuated the (continuous) symmetries of the underlying background structures. Secondly we modified the picture of the configuration space of the theory as a principal fiber bundle structure whose fibers are the ``gauge orbits'' generated by the symmetry group and the base space is the real fundamental configuration space. Thirdly, we defined (i) a notion of coordinates shifting according to the action of the elements of the symmetry group, (ii) a connection over the fiber bundle that specifies a covariant derivative, and (iii) a flat ``kinetic'' metric given in terms of shifted coordinates and covariant derivatives, which is curved by a conformal factor depending on potential $V$ of the system. Finally, we used this machinery to set a variational problem of the Jacobi-type in order to get the equations of motion. Given that BM is formulated over $\mathbb{R}^{3N}$, the first two steps are implementable as in the previous case. The real problem is setting a variational problem (\ref{jac}) for this theory, especially with respect to finding an appropriate conformal factor $\mathcal{C}$ that accounts for the inter-particle ``influences''. This is obvious because BM is \emph{not} a classical theory in that the behavior of the particles in a configuration is not only dictated by the physical interactions encoded in a classical potential $V$, but also - and more importantly - by the structure of the wave function. It is in fact $\Psi$ that confers the non-local behavior on a configuration of particles, with the result that a Bohmian configuration is much more ``rigid'' than a classical one. Does this fact spoil the above considered framework? Perhaps not, provided we ``cheat'' a little bit. The cheating we are referring to consists in camouflaging BM as a ``pseudo-Newtonian'' theory.
This can be easily done by writing the wave function in polar form (i.e. by putting $\Psi(\mathbf{Q},t)=R(\mathbf{Q},t)e^{iS(\mathbf{Q},t)}$, $R$ and $S$ being real functions), and then inserting it into (\ref{sch}). The resulting formula will consist of two coupled equations, one expressing the conservation of $|\Psi|^{2}$ along particles' trajectories, the other being:
\begin{equation}\label{HJ}
\frac{\partial S}{\partial t}+\sum_{i=1}^{N}\frac{(\nabla_{i}S)^{2}}{2m_{i}}+V+\mathcal{V}=0.
\end{equation}
Since our starting theory is (\ref{bm}), that is a non-relativistic theory of $N$ point-like particles, what we have done is basically to rewrite it as a (quantum) Hamilton-Jacobi theory, where the $k$-th particle's velocity is $\frac{\boldsymbol{\nabla}_{k}S}{m_{k}}$ (which is nothing but (\ref{gui2}) written in polar form), and the total potential has now another ``quantum'' term besides the classical one $V$, namely:
\begin{equation}\label{qpot}
\mathcal{V}=-\sum_{i=1}^{N}\frac{1}{2m_{i}}\frac{\nabla_{i}^{2}R}{R}.
\end{equation}
The reason why we call it a potential becomes manifest if we derive (\ref{HJ}) with respect of $\boldsymbol{\nabla}_{k}$, thus arriving at a ``Newtonian-like'' equation of motion for the $k$-th particle, which reads:
\begin{equation}\label{bmgnutto}
m_{k}\frac{d^{2}\mathbf{q}_{k}}{dt^{2}}=-\boldsymbol{\nabla}_{k}\big(V+\mathcal{V}\big).\footnote{It is possible to arrive at the same expression by differentiating (\ref{gui}) with respect to time, which stresses the fact that we are just fiddling with the formalism of our starting theory (\ref{bm}).}
\end{equation}
Clearly, (\ref{bmgnutto}) resembles (\ref{gnutto}) with the additional term (\ref{qpot}) being a ``quantum'' potential. It is very important to stress the fact that such a rewriting of the theory is just a convenient disguise totally deprived of any substantial physical meaning.\footnote{At least, in the present approach. In Bohm's own formulation the quantum potential was regarded as primary, and there exists also a ``quantum force'' formulation of the de Broglie-Bohm theory where (\ref{bmgnutto}) is in fact the fundamental equation (see \citealp{412} for an introduction to this latter approach, and see also \citealp{411}, sections 1.4, 1.5, and 1.6, for a discussion of the utility of (\ref{bmgnutto}) in BM).} The theory (\ref{bm}) is, in fact, a first order theory where accelerations - let alone ``quantum'' forces - play no fundamental role at all. Furthermore, as already stressed in section \ref{2}, one key feature of (\ref{bm}) is that it delivers us from the need of postulating the existence of $\Psi$ as a physical field that exerts a ``quantum'' force to the particles. Hence, the purpose of calling the attention on (\ref{HJ}) and (\ref{bmgnutto}) resides only in the fact that (\ref{bm}) is not \emph{structurally} incompatible with the best-matching framework sketched above. This claim becomes straightforward once we realize that, by interpreting (\ref{HJ}) as a quantum Hamilton-Jacobi equation, the quantum behavior of the particles can be accounted for by a modified best-matching principle involving an action of the form (\ref{jac}), which features a conformal factor $\mathcal{C}=\mathcal{E}-\mathcal{W}$ - where $\mathcal{E}$ is the ``total energy'' of the particle system,\footnote{That is, under the parameter fixing (\ref{gnutfix}), we expect to find $\mathcal{E}=-\frac{\partial S}{\partial t}$.} and $\mathcal{W}=(V+\mathcal{V})$ is the total potential -, and a ``kinetic'' metric $T$ depending on $\sum_{i}(\nabla_{i}S)^{2}$. If such a theory could be consistently worked out, then the quantum information encoded in the wave function would be reduced to geometrical features of the Riemannian structure defined over shape space.\\
However, showing that BM can be put in a form that looks classical is not enough to prove that it can be stripped of its Newtonian background using Barbour's strategy. This is because, in this case, both the kinetic term and the additional quantum potential are not interpretable using classical notions: they are rather an elaborated way to disguise the wave function. Indeed, in general, the kinetic term has a non-trivial dependence on the wave function's phase $S$, while the potential (\ref{qpot}) is not a preassigned function depending on particles' positions only, but has a complicated structure involving a spatial gradient of the wave function's amplitude. Therefore, setting up a best-matching principle of the form (\ref{jac}) involving a kinetic term and a total potential as they appear in (\ref{HJ}) requires a certain amount of technical work, including (i) specifying the additional symmetries that these quantities have, (ii) figuring out how to properly parametrize them (since both $S$ and $R$ are in general time-dependent functions), (iii) spelling out a rigorous mathematical procedure to perform a variation on them, and (iv) considering how putting these terms in (\ref{jac}) changes the mathematical picture.\\
To summarize: given that from a best-matched theory (\ref{mot}) it is possible to recover a theory whose equations of motion have the form (\ref{gnutto}), and given that the equations of motion of BM can be massaged into such a form, we conclude that there is no a priori or self-evident argument that speaks against the possibility of considering BM as being reducible to a more fundamental best-matched theory. Therefore, since best-matching can be seen as an effective strategy to strip a theory of its background spatiotemporal structures in a non-trivial sense, we have shown that there are good reasons to think that BM can be made background independent.

\section{Conclusion}
The heuristic argument put forward at the end of the previous section on why the best-matching strategy should probably work in BM rests on a rather simple observation: the dynamics of Bohmian particle systems is perfectly embedded in Newtonian spacetime as the standard classical dynamics is. This means that both in a Bohmian and in a Newtonian setting we have swarms of particles moving in Euclidean $3$-space under the ticking of a universal immutable clock. What changes in the two pictures is the behavior of the particles - and, indeed, such a difference is extremely radical, being at the root of the classical/quantum divide. Nonetheless, with respect to considerations regarding the symmetries of spacetime, it seems that we can safely assume that a Bohmian system has a well-defined kinetic energy (although this is not a fundamental quantity, but supervenes on the wave function's phase), and is subjected to ``yet another potential'' (although a rather strange and complicated one). In turn, this hints at the fact that we can absorb these quantities in the key ingredient of a best-matched theory, namely the conformally deformed metric $\mathcal{C}T$ on shape space, which accounts for the theory's dynamics through the implementation of the best-matching action principle. Why is the fact that BM is embedded in Newtonian spacetime so important? Because, as long as the main result of the best-matching procedure is eliminating the background spatiotemporal structures from the picture, the fact that BM and Newtonian mechanics share the same spatiotemporal arena assures us that best-matching does not care much about how much "strange" particles' movements in such an arena are, but just cares about the arena itself. Moreover, the ``rigidity'' of Bohmian configurations, due to the fact that the quantum potential is independent on the inter-particle separations - so that the behavior of each particle strongly depends on the behavior of all the others -, naturally fits into the ``holistic'' perspective of best-matching, since this framework needs to focus on universal configurations in order to recover a useful notion of ``universal clock''. From this point of view, BM is even more akin to best-matching than classical mechanics, where the fact that the standard potentials (e.g. the gravitational one) fall off with the inverse of the inter-particle separation renders each particle's behavior actually dependent just on the behavior of its neighbors, thus making the talk of configurations as a whole less necessary.\\
To conclude, even if we have not provided a final answer to the question that entitles this paper, we have nonetheless reached two important intermediate results, namely, (i) we have pointed out a promising strategy to render BM background independent (best-matching), and (ii) we have shown that BM is not structurally incompatible with such a strategy. The next step will be to show beyond heuristics that BM can be made background independent by implementing the best-matching procedure as suggested in the previous section. With no doubt, this will be quite a formidable task for mathematicians and physicists but, as this paper shows, the route is open and ready to be walked.

\pdfbookmark[1]{Acknowledgements}{acknowledgements}
\begin{center}
\textbf{Acknowledgements}:
\end{center}
I am grateful to an anonymous referee for helpful comments on an earlier draft of this paper. Research contributing to this paper was funded by the Swiss National Science Foundation, Grant no. $105212\_149650$.

\pdfbookmark[1]{References}{references}
\bibliography{biblio}
\end{document}